\documentclass[11pt,a4paper]{article}
\usepackage{jheppub}

\title{A Challenge to Entropic Gravity}

\author{Jonathan J. Roveto,}
\author{Gerardo Mu\~noz}

\affiliation{Department of Physics, California State University - Fresno, \\
2345 E. San Ramon Ave., Fresno, USA}

\emailAdd{jroveto@csufresno.edu}
\emailAdd{gerardom@csufresno.edu}

\abstract{In a recent publication in this journal, Erik Verlinde attempts to show that gravity should be viewed not as a fundamental force, but rather as an emergent thermodynamic phenomenon arising from an unspecified microscopic theory via equipartition and holography. This paper presents a challenge to his reformulation of gravity. A detailed examination of Verlinde's derivation leads to a number of questions that severely weaken the claim that such a theory correctly reproduces Newton's laws or Einstein gravity. In particular, we find that neither Newtonian gravity nor the Einstein equations are uniquely determined using Verlinde's postulates.}


\notoc
\begin{document}

\maketitle

\newpage
\section{Introduction}

Using a polymer-based model, thermodynamics, and a specific interpretation of the holographic principle, Verlinde \cite{Verlinde} claims to have shown that gravity is a force solely caused by an exchange of information on a holographic screen. In essence, he concludes that gravity is not a fundamental force, but the result of changes in entropy of an unknown microscopic theory. Our purpose in this paper is to examine the proposed entropic origin of gravity in detail. In particular, we investigate whether Verlinde's claim that his postulates are sufficient to derive Newtonian gravity and the Einstein field equations can be upheld. Although it does not contradict the general idea of a relationship between thermodynamics and gravity, our analysis shows that neither Newtonian gravity nor the Einstein equations can be unambiguously reproduced by this particular treatment of the thermodynamics/gravity connection.

\section{Thermodynamics and Gravity}

The original motivation for treating gravity thermodynamically may be traced to Bekenstein's observation \cite{Bekenstein1973,Bekenstein1974} that the areas of black hole horizons are strongly analogous to entropy in some of their properties or, more precisely, to his conjecture that entropy and horizon area are proportional. The Bekenstein-Hawking entropy of a black hole system is  (we employ units where $c=\hbar=k_B=1$ but we will leave $G$ in place)
\begin{eqnarray}
\label{3}
S = {{A}\over{4 G}}
\end{eqnarray}
where $A$ is the area of the black hole event horizon. Bekenstein's initial proposal \cite{Bekenstein1973} was based on Hawking's area theorem \cite{Hawking1971}, and the connection was bolstered by the laws of black hole mechanics developed by Bardeen, Carter and Hawking \cite{BCH1973}. But the analogy did not acquire true physical significance until Hawking's discovery \cite{Hawking1975} that quantum mechanical effects allow black holes to radiate with a thermal spectrum at a temperature given by
\begin{eqnarray}
\label{1}
T_{Hawking}={{\kappa}\over{2\pi}}
\end{eqnarray}
where $\kappa$ is the surface gravity of a near-horizon observer. For our present purposes, another essential fact was provided by Unruh \cite{Unruh}. The Unruh effect tells us that an accelerating observer sees itself surrounded by a heat bath whose temperature is given by
\begin{eqnarray}
\label{2}
T_{Unruh}={{a}\over{2\pi}}
\end{eqnarray}
where $a$ is the proper acceleration of the observer. The resemblance to (\ref{1}) is apparent.

\section{Gravity as an Entropic Force}

We will now review the details of Verlinde's proposal that gravity can be described using the holographic principle. He begins with the premise that the entropy contained within a region of spacetime can be mapped to an appropriately chosen holographic screen. On one side of the screen, he assumes, the spacetime in the usual sense has emerged, and macroscopic variables such as positions and coordinates may be employed. On the other side, the physics is encoded in a microscopic theory as yet unknown to us. Motivated by an argument due to Bekenstein \cite{Bekenstein1973}, Verlinde postulates that a particle approaching this holographic screen causes the entropy on the screen to change by $\Delta S = 2\pi$ when it comes within a Compton wavelength, $\Delta x = m^{-1}$, and writes this as
\begin{eqnarray}
\label{4}
\Delta S = 2\pi m \Delta x.
\end{eqnarray}

The analogy of osmosis across a semi-permeable membrane is used to explain how a particle would have a reason to cross this holographic screen. If there is an entropically favorable configuration available by moving towards the screen, the particle would do so. He reasonably posits that such a force is not fundamental; it is rather an effective force due to the microscopic theory involved. This effective force is given by
\begin{eqnarray}
\label{5}
F \Delta x = T \Delta S.
\end{eqnarray}
Invoking the Unruh temperature (\ref{2}) in conjunction with (\ref{4}), he finds Newton's 2nd Law
\begin{eqnarray}
\label{6}
F=ma.
\end{eqnarray}

The attempt to derive Newton's law of universal gravitation proceeds along similar lines. He postulates that the number of ``bits" on the holographic screen is given by 
\begin{eqnarray}
\label{7}
N={{A}\over{G}}
\end{eqnarray}
where $A$ is the area of the holographic screen. This screen encloses a mass, $M$, with energy
\begin{eqnarray}
\label{8}
E=M.
\end{eqnarray}
If the bits associated with this energy are distributed on the holographic screen according to equipartition, we have
\begin{eqnarray}
\label{9}
E={{1}\over{2}}NT.
\end{eqnarray}
Finally, using (\ref{4}), (\ref{5}), and (\ref{7} - \ref{9}), along with the fact that the area of the holographic screen is $4\pi R^2$, one finds Newton's law of gravity,
\begin{eqnarray}
\label{10}
F={{GMm}\over{R^2}}\, .
\end{eqnarray}

Verlinde then proceeds to a derivation of the Poisson equation for Newtonian gravity. Since ${\bf a} = -\nabla \Phi$, one can do a few simple substitutions to write the entropy change in terms of $\Phi$, the Newtonian potential that Verlinde argues simply keeps track of the ``information" of the system, as
\begin{eqnarray}
\label{11}
{{\Delta S}\over{N}} = -{{\Delta \Phi}\over{2}}.
\end{eqnarray}
Consider a static matter distribution $\rho$ completely contained within a volume $V$, and enclosed by a holographic screen $S = \partial V$ specified by $\Phi = {\rm constant}$. The temperature may be written in terms of the normal derivative of the potential,
\begin{eqnarray}
\label{12}
T={1\over{2\pi}} \, {\bf \hat n} \cdot {\nabla \Phi}.
\end{eqnarray}
with ${\bf \hat n}$ the outward normal to the holographic screen enclosing the static mass distribution. 
The density of these bits on the holographic screen $S $ is postulated to be
\begin{eqnarray}
\label{13}
dN = {{1}\over{G}}\, dA.
\end{eqnarray}
Equipartition tells us that the energy $E$ associated with these bits is
\begin{eqnarray}
\label{14}
E={{1}\over{2}}\int_{\partial V} TdN.
\end{eqnarray}
Using holography to set $E=M$, with $M$ the total mass within $V$, together with (\ref{12}) - (\ref{14}), we get
\begin{eqnarray}
\label{15}
M={{1}\over{4\pi G}}\int_{\partial V} \nabla \Phi \cdot d\bf{A}.
\end{eqnarray}
Expressing $M$ as an integral of the mass density $\rho$ over the volume enclosed, Verlinde claims that (\ref{15}) can hold in general only if the potential obeys
\begin{eqnarray}
\label{16}
\nabla^2 \Phi = 4\pi G \rho,
\end{eqnarray}
the Poisson equation for Newtonian gravity.

This treatment of gravity can now be extended to a relativistic setting. Following Wald \cite{Wald}, one can show that the acceleration of a stationary observer in a static curved space-time can be written as
\begin{eqnarray}
\label{17}
a^b = e^{-2\phi}\xi^a \nabla_a \xi^b
\end{eqnarray}
where $\xi^b$ is the timelike Killing vector and 
\begin{eqnarray}
\label{18}
\phi = {{1}\over{2}}\ln (-\xi^a \xi_a)
\end{eqnarray}
is the natural generalization of the Newtonian potential. In terms of $\phi$, the acceleration is given by
\begin{eqnarray}
\label{19}
a^b = -\nabla^b\phi .
\end{eqnarray}
Following the same procedure as in the non-relativistic case we find that the temperature may be written as
\begin{eqnarray}
\label{20}
T={e^{\phi}\over{2\pi}}N^b\nabla_b\phi
\end{eqnarray}
where the appropriate redshift factor is in place and $N^b$ is a spacelike unit outward normal perpendicular to both the holographic screen $S$ defined by $\phi =$ constant, and to $\xi^a$. The entropy gradient is rewritten
\begin{eqnarray}
\label{21}
\nabla_a S = -2\pi m\nabla_a \phi .
\end{eqnarray}
Using the same procedure as for Newtonian gravity, Verlinde shows that
\begin{eqnarray}
\label{22}
M={{1}\over{4\pi G}}\int_S e^{\phi} \nabla_b \phi \, N^b dA,
\end{eqnarray}
Stokes' theorem and the identity $\nabla^b \nabla_b \xi_a = - R_{ab}\xi^{b}$ produce the alternative form
\begin{eqnarray}
\label{24}
M={{1}\over{4\pi G}}\int_{\Sigma}R_{ab} n^a \xi^b dV
\end{eqnarray}
where $\Sigma$ is a three-dimensional volume with boundary $\partial \Sigma = S$, and $n^a$ is normal to $\Sigma$. One would naturally expect the left hand side to be given by an integral over $\Sigma$ of the energy-momentum tensor. By comparing the properties of the integrands, Verlinde argues that the correct combination is
\begin{eqnarray}
\label{25}
2\int_{\Sigma}(T_{ab}-{{1}\over{2}}Tg_{ab}) \, n^a \xi^b dV = {{1}\over{4\pi G}}\int_{\Sigma} R_{ab}n^{a}\xi^{b}dV.
\end{eqnarray}
The claim is that this derivation is sufficient to show that Einstein's equations follow from treating gravity as an emergent phenomenon.

\section{Challenging Entropic Gravity}

In this section we present some obstructions that arise in Verlinde's approach to gravity as an entropic force. Eq. (\ref{15}) is the basis for our challenge to this theory in the Newtonian case. Using Gauss' law and writing the mass $M$ as an integral of the mass density $\rho$ over the volume enclosed by the equipotential surface, we obtain
\begin{eqnarray}
\label{26}
\int_V \rho dV = {{1}\over{4\pi G}}\int_V \nabla^2 \Phi dV.
\end{eqnarray}
Verlinde's central claim is that the Poisson equation follows from equating the integrands. Recall, however, the fundamental assumption that $V$ is a volume bounded by a holographic screen specified by $\Phi =$ constant. This implies that any modification to the Poisson equation of the form
\begin{eqnarray}
\label{28}
\nabla^2 \Phi + {\bf u}\cdot \nabla f(\Phi) = 4\pi G \rho,
\end{eqnarray}
where $\nabla \cdot {\bf u} = 0$ and $f(\Phi)$ is an arbitrary function of the potential, will also satisfy (\ref{26}) at the integral level. The proof is trivial:
\begin{eqnarray}
\label{29}
\int_V \left[ \nabla^2 \Phi + {\bf u} \cdot \nabla f(\Phi) \right]dV &=& \int_V \left[ \nabla^2 \Phi + \nabla \cdot \left({\bf u} f(\Phi)\right) \right] dV \\ \nonumber
&=& \int_V \nabla^2 \Phi dV + f(\Phi) \int_{\partial V} {\bf u} \cdot d {\bf A} \\ \nonumber
&=& \int_V \nabla^2 \Phi dV \\ \nonumber
\end{eqnarray}
where we have used the fact that $f(\Phi) $ is constant on the equipotential surface bounding $V$, and that the integral over a closed surface of any divergence-free vector field must vanish. Note that, in general, ${\bf u}$ introduces nonlinear, higher-derivative terms as well as a breakdown of homogeneity and isotropy in empty space. It follows from this simple counterexample that the basic postulates of the theory are incapable of reproducing the Poisson equation for Newtonian gravity. In particular, we recognize the appearance of unacceptable alternatives of the form (\ref{28}) as directly related to the implementation of holography in the theory.

Our challenge to the relativistic derivation runs along similar lines. Starting with equation (\ref{24}) and accepting as valid (for the moment - see below) the argument leading to (\ref{25}) we have
\begin{eqnarray}
\label{38}
\int_{\Sigma}(\Theta _{ab} \, \xi^b) \,n^a dV = 0
\end{eqnarray}
where $\Theta_{ab}= R_{ab}  - 8\pi G ( T_{ab} - {{1}\over{2}}T g_{ab})$. Verlinde points out that the normal $n^a$ in (\ref{38}) is arbitrary, and suggests adapting Jacobson's \cite{Jacobson} local argument for $\xi^b$ to show that one may uniquely determine all components of the Einstein equations at the integrand level. However, as a consequence of the Killing equation $\nabla_a \xi_b + \nabla_b \xi_a =0$, the normal to the holographic screen $\nabla_b \phi$ is everywhere orthogonal to $\xi^b$. Hence, even at a local level, the field equations
\begin{eqnarray}
\label{39}
R_{ab} -8\pi G ( T_{ab} - \tfrac{1}{2} T g_{ab}) + f(R,T, \phi )\nabla_aU(\phi) \nabla_b V(\phi) = 0
\end{eqnarray}
with $f(R,T, \phi )$ an arbitrary function of the Riemann tensor, the energy-momentum tensor and $\phi$, and $U(\phi)$ and $V(\phi)$ arbitrary functions of $\phi$, will yield the same answer as $\Theta_{ab}$ when contracted with $\xi^b$. Clearly then, the integral relationship (\ref{38}) holds for (\ref{39}) as well as for the Einstein equations. Note that in general relativity adding terms such as $f\nabla_aU \nabla_b V$ would not be acceptable. But because the holographic screens play a fundamental physical role in Verlinde's theory, such terms cannot be dismissed off hand in his approach to entropic gravity. 

Furthermore, the two-dimensional screen $S$ admits two vectors, $e^{\alpha}_a$, parallel to the surface. Since $e^{\alpha}_a \xi^a = 0$, one could add terms such as $\sum\limits_{\alpha , \beta = 1}^{2} C_{\alpha \beta} \, e^{\alpha}_a e^{\beta}_b$, where $C = C(R, T, \phi )$, without altering equation (\ref{38}).

\section{Discussion}

We have presented some challenges to Verlinde's theory of entropic gravity. Quite obviously, these challenges do not purport to undermine the possible origin of gravity as the thermodynamic limit of a more fundamental theory. Our comments simply show that proclaiming the end of gravity as a fundamental force based on Verlinde's approach is premature.

While the objections of the previous section are fairly elementary, there are other issues that should be addressed before entropic gravity can be regarded as providing deeper insights into the nature of spacetime and gravitation than conventional general relativity. For instance, Verlinde's statement that the left hand side of (\ref{25}) can be fixed by comparing the properties of the integrands on both sides is unwarranted since the problem is precisely to try to determine the properties of these integrands based on the integral relations implied by the theory. If one insists on reasoning at the level of the integrands, a theorem by Lovelock \cite{Lovelock} almost immediately and unambiguously produces the Einstein equations without the need for thermodynamic input.

One might also ask why equipartition, a classical concept, can (or should) be combined with the Unruh temperature, a quantum field theory concept, to arrive at classical laws such as Newtonian gravity or general relativity without so much as a hint of any quantum corrections. This is clearly disappointing, since much of the appeal of the thermodynamic approach is tied to the insights it may provide on the putative microscopic degrees of freedom. But it may also point to a serious flaw in Verlinde's assumptions. Indeed, both in Newtonian gravity and in general relativity it is easy to construct exact solutions for which $T =0$, as calculated from (\ref{12}) or (\ref{20}), in clear violation of the third law of thermodynamics.  In the absence of quantum corrections, this violation is an exact prediction of Verlinde's theory. Furthermore, Hossenfelder has argued \cite{Hossenfelder} that one need not invoke the Unruh observer nor the holographic principle in the first place to derive a boundary theory of Newtonian gravity. This is probably a good thing considering others \cite{Singleton} have shown that non-negligible changes to $F=ma$ can be found if one considers an observer undergoing centripetal acceleration. 

The derivation of the Einstein equations gives rise to some additional concerns. One might ask on what grounds can we justify modifying (\ref{38}) with functions involving $\phi$. The potential, $\phi$, be it in the equipotential surfaces of Newtonian gravity or the constant redshift surfaces of relativistic gravity, are abstract concepts. In Verlinde's theory, these surfaces are promoted to actual physical objects - bookkeepers of ``information". The postulate of holography makes them no less fundamental to this theory than the Riemann or energy-momentum tensors. But even if we reconcile ourselves to the new status of $\phi$, one might further ask ``why Einstein?" Verlinde's derivation relied on the entropy-area law of Bekenstein. However, as Visser \cite{Visser} has shown using Euclidean techniques, the entropy-area law is not unique to Einstein's general relativity but holds also for (at least) fairly general (Riemann)${}^2$ Lagrangians in four dimensions. It is therefore difficult to understand why postulating the entropy-area connection should lead in a unique manner to the Einstein equations.

Finally, one may wonder why the cosmological constant fails to appear. Jacobson's argument \cite{Jacobson} based on null horizons incorporates the cosmological constant in a natural way. There does not seem to be a simple way of introducing a cosmological constant in Verlinde's theory.

\end{document}